\documentclass[prl,twocolumn,showpacs,amsmath,amssymb]{revtex4}

\usepackage{graphicx}
\usepackage{dcolumn}
\usepackage{bm}

\begin{document}

\preprint{APS/123-QED}

\title{Escape from a zero current state in a one dimensional array of Josephson junctions}

\author{K.~Andersson}
\email{karina@nanophys.kth.se}
\author{D.~B.~Haviland}
\affiliation{Nanostructure physics, KTH\\
AlbaNova universitetscentrum\\
S-106 91 Stockholm, SWEDEN }

\date{\today}

\begin{abstract}
A long one dimensional array of small Josephson junctions exhibits Coulomb blockade of Cooper pair 
tunneling. This zero current state exists up to a switching voltage, $V_{sw}$, where there is a 
sudden onset of current. In this paper we present histograms showing how $V_{sw}$ changes with 
temperature for a long array and calculations of the corresponding escape rates. Our analysis of 
the problem is based on the existence of a voltage dependent energy barrier and we do not make any 
assumptions about its shape. The data divides up into two temperature regimes, the higher of which 
can be explained with Kramers thermal escape model. At low temperatures the escape becomes 
independent of temperature.
\end{abstract}

\maketitle

Escape from a metastable state can be used to describe a wide variety of physical phenomenon, from 
solid state diffusion \cite{diFoggio:diffusionHydro:82} to magnetic switching 
\cite{Sun:magnSwitch:01}, and has for a long time attracted attention from both a theoretical and 
an experimental point of view (for a review see e.g. \cite{hanggi:50yrs:90}). Thermal escape over 
an energy barrier was studied theoretically by Kramers in the 1940s \cite{kramers:escape:40}. More 
recently, the influence of quantum tunneling on the escape process has been an important means of 
studying the quantum behavior of macroscopic variables. In this context, escape from the zero 
voltage state of a Josephson junction has been thoroughly studied 
\cite{martinis:phasedifference:87, devoret:QuantumEffectRev:92}. In this paper we adopt the Kramers 
model to analyze the escape from the zero \emph{current} state of a one dimensional array of 
Josephson junctions exhibiting a Coulomb blockade of Cooper pair tunneling. We find that the escape 
process is well described by thermal activation at higher temperatures, but at lower temperatures 
the escape process becomes independent of temperature.

In the Kramers thermal activation model, the rate of escape $\Gamma$, over an energy barrier 
$\Delta U$, is given by, 
\begin{equation}
  \Gamma=\kappa{\frac{\omega _p}{2 \pi}}\exp({-{\frac{\Delta U}{k_BT}}})
  \label{eq:ThermEsc}
\end{equation}
where $k_B$ is Boltzmann's constant and $\omega _p$ is the attempt frequency, or the frequency of 
small oscillations around the metastable minimum. The dimensionless transmission factor $\kappa$ 
depends on the shape of the barrier and the damping \cite{hanggi:50yrs:90}. This model has been 
applied to single Josephson junctions, where the energy barrier is a one dimensional tilted 
washboard potential, and the tilt is controlled by bias current applied to the junction, $\Delta 
U(I)$ \cite{fulton:swCurr:74,martinis:phasedifference:87,devoret:QuantumEffectRev:92}. Long 
Josephson junctions (modeled by parallel arrays of single junctions) have also been studied, and in 
this case the escape is from a metastable minimum in a multi-dimensional space 
\cite{castellano:escLongJJ:96}. All Josephson junction systems studied with this technique thus far 
exhibit a zero voltage state with a finite supercurrent, above which the system switches to a 
finite voltage state. A voltage across the junction develops when the superconducting phase, 
$\phi$, across the junction escapes over the Josephson potential energy barrier to a free-running 
state ($V\propto \dot{\phi}$). The escape can be due to thermal activation or quantum tunneling of 
the junction phase through the energy barrier. The important quantities for the escape problem are 
the height of the Josephson potential and the attempt frequency, which are given in terms of other 
junction parameters such as normal state resistance, superconducting energy gap and junction 
capacitance. This escape description applied to the Josephson junction is based on the classical 
dynamics of the Josephson phase, as embodied in the Josephson effect. 

In this letter we analyze the Coulomb blockade of Cooper pair tunneling, a phenomena which is in 
many ways dual to the Josephson effect \cite{agren:kinInd:01, agren:licThesis:00}. In 
one-dimensional arrays of small capacitance Josephson junctions, the coupling of the Josephson 
phase to dissipation can be manipulated so that there are very large quantum fluctuations of the 
phase \cite{haviland:SItrans:00}. In this case, the number of Cooper pairs becomes a classical 
variable, and transport through the array is described by a Coulomb blockade, where a zero current 
state persists until a switching voltage is exceeded. One- and two-dimensional arrays of small 
capacitance Josephson junctions have been well studied in the context of the 
superconductor-insulator transition \cite{fazio:phaseTrans:01}. Here we focus on analysis of 
transport properties of a 1D array deep in the insulating regime. We do this by studying 
fluctuations of the switching voltage and analyizing them in the context of the Kramers model. This 
is the first time to our knowledge that such an experimental study has been carried out on Cooper 
pair tunneling in the insulating regime.

The array is made of Al, with Al$_2$O$_3$ tunnel barriers, on a SiO$_2$ surface using the standard 
two angle evaporation technique \cite{andersson:licThesis:99}. Electron beam lithography was used 
to define the mask pattern, which made 400 junctions in series. Each junction had dimensions 
$100$nm$ \times 150$nm, and each island was connected by two junctions in parallel, forming a 
Superconducting QUantum Interference Device (SQUID). The charging energy $E_C$ of the SQUID was 
$E_C\equiv e^2/2C=59\mu$eV, where we have assumed a specific capacitance of 
$45fF/\mu$m$^2$\cite{delsing:arraysprb:94}. The junctions normal state resistance $R_n$ controls 
the Josephson energy at zero magnetic field, $E_{J0}=(R_Q/R_n) \Delta/2$ where $\Delta=200 \mu$eV 
is the superconducting energy gap of Al. For the array presented here we had $R_n=4.6$k$\Omega$ per 
SQUID, or $E_{J0}=142\mu$eV. 

The advantage of the SQUID geometry is that we can tune the Josephson coupling by applying a 
magnetic flux to the SQUID loops $E_J=E_{J0}\mid \cos(\pi BA/\Phi_0)\mid$, where $B$ is the 
magnetic field perpendicular to the SQUID loop, $A$ is the effective area of the loop (in this case 
$0.85\times 0.2 \mu $m$^2$) and $\Phi _0=h/2e=2.06\times 10^{-15}$Wb is the superconducting flux 
quantum. 

The arrays were measured in a dilution refrigerator with well filtered measurement lines. The chip 
is mounted on a socket which is in turn mounted in a RF tight copper can. The leads into the can 
are 1 meter of Thermocoax which attenuates at high frequency \cite{zorin:thermocoax:95}. The can 
and filtered leads are mounted at the lowest temperature. The bonding pads and contact leads on the 
chip were fabricated of Au to within $50\mu$m of the array edge. Low noise and high input impedance 
amplifiers were used to measure the voltage and current. 

\begin{figure}[b]
\center 
\includegraphics[width=0.9\columnwidth]{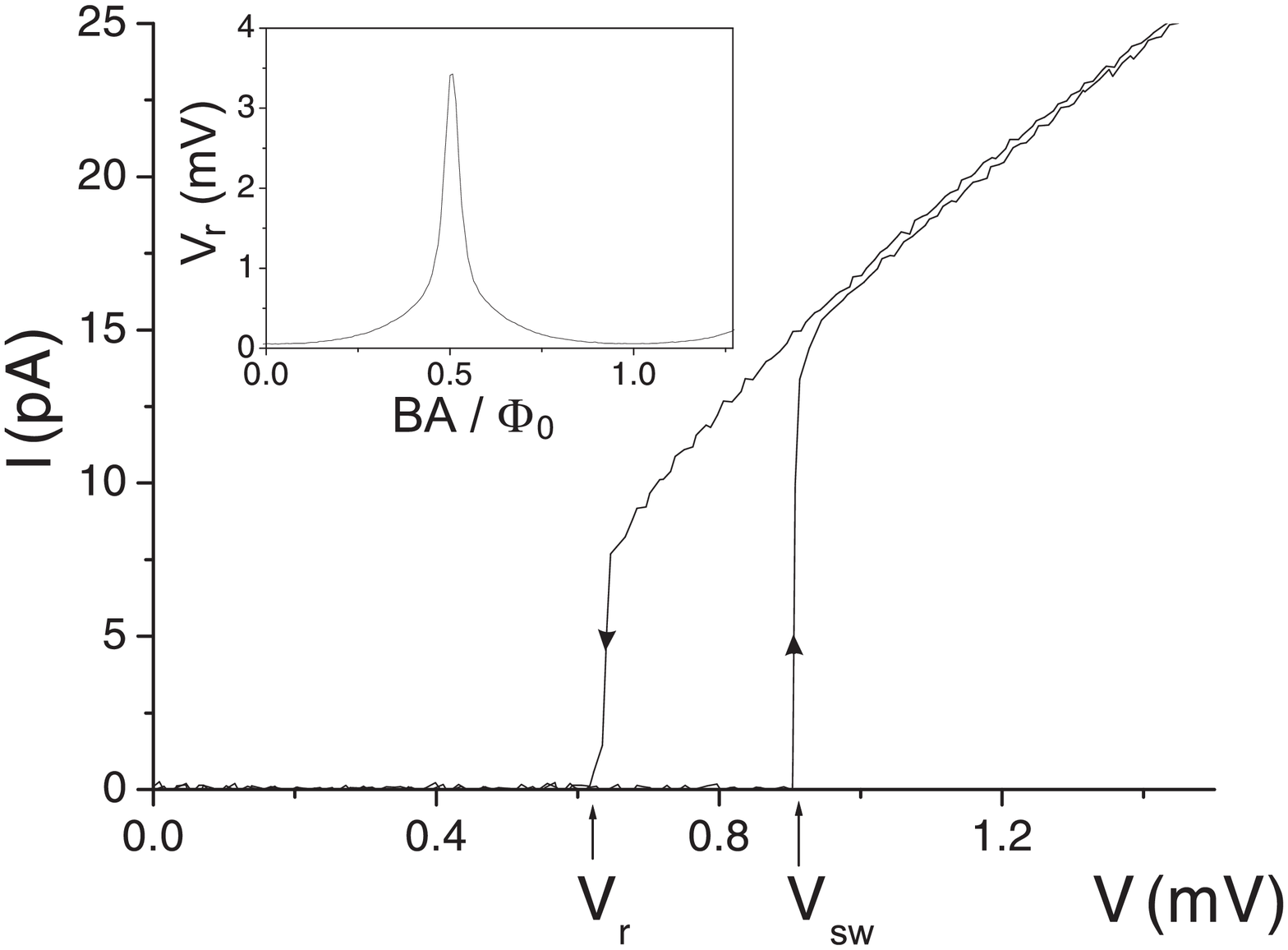} 
\caption{\label{fig:CBCPT}Current voltage characteristics of a typical long one dimensional array 
of Josephson junctions. Voltage bias gives a hysteretic $I-V$ curve, with a switching voltage, 
$V_{sw}$, and the retrapping voltage, $V_r$. The arrows on the curve indicate the direction of the 
sweep. The inset shows how $V_r$ is modulated with B.} 
\end{figure}

In Fig.\ \ref{fig:CBCPT} we see a typical current-voltage (I-V) characteristic of a 1D array in the 
Coulomb blockade state. When increasing the bias voltage across the array the Coulomb blockade 
state is interrupted by a sudden onset of current at the switching voltage, $V_{sw}$. Subsequently 
lowering the bias voltage across the array, the Cooper pairs are retrapped at a voltage $V_r < 
V_{sw}$, giving rise to a hysteresis in the current-voltage characteristics. The retrapping voltage 
remains the same at fixed magnetic field, whereas the switching voltage fluctuates. Both $V_{sw}$ 
and $V_r$ depend on $E_J$ and are modulated with magnetic field having a period associated with one 
flux quantum in each SQUID loop, as can be seen in the inset of Fig.\ \ref{fig:CBCPT}. This 
modulation is clear evidence that the switching voltage is associated with the Cooper pair 
tunneling in the array.

By making $10^4$ sweeps of the bias voltage and each time registering the switching voltage 
$V_{sw}$, we obtain a switching histogram representing the probability distribution $P(V)$ from 
which the escape rate can be calculated through the formula \cite{fulton:swCurr:74},
\begin{equation}
  \Gamma(V)={\frac {1}{\Delta V}}{\frac{dV}{dt}}\ln{\left({\sum_{v\geq 
V}P(v)\text{\Huge/}{\sum_{v\geq V+\Delta V}P(v)}}\right)}
\label{eq:CalcG}
\end{equation}
where $dV/dt$ is the ramp rate of the voltage and $\Delta V$ is the width of the bins in the 
histogram.

To establish that the switching process is indeed sensitive to the applied voltage, we have varied 
the sweep rate by changing the frequency of our triangular sweep between $5$ and $200$Hz, while 
keeping the amplitude of the sweep constant at $1.2$mV. For lower sweep rates the histogram shifted 
to lower voltages as we are less likely to reach higher voltages before escape occurs. However, 
when the escape rate $\Gamma(V)$ was the calculated according to (\ref{eq:CalcG}) and comparison 
was made for different sweep rates, we found see that for sweep rates lower than $0.12$V/s, the 
escape rate at a given voltage was essentially unaffected by the sweep rate. We also found that for 
sweep rates higher than $0.12$V/s, the escape rate $\Gamma(V)$ becomes sensitive to the sweep rate. 
This limitation on sweep rate was set by the biasing and current measurement circuit together with 
the stray capacitance of the cryostat leads. From this analysis we conclude that the switching 
process, when measured with sweep speed below $0.12$V/s, is truly a voltage dependent escape 
problem. We also investigated if the switching process is affected by the time spent in the 
dissipative (finite current) state by putting in a delay and waiting at zero voltage between 
successive ramps. The result of this investigation was that the switching histogram for fixed sweep 
rate and temperature remained the same, independent of delay time. 

\begin{figure}[t]
\center 
\includegraphics[width=0.9\columnwidth]{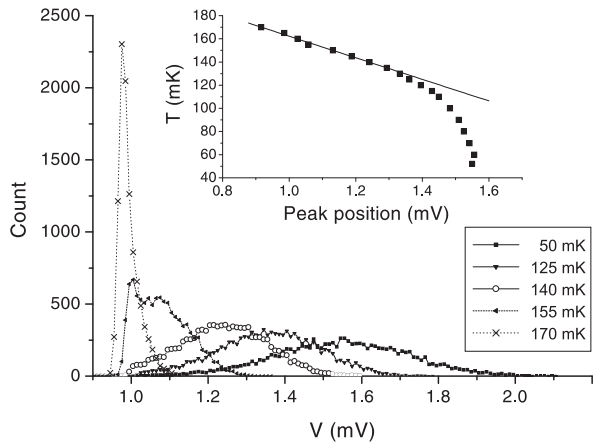} 
\caption{\label{fig:Hist}A selection of histograms containing $10^4$ switching events, each at 
different temperatures. As the temperature increases the histogram shifts towards lower voltages. 
At high temperatures it is evident how the retrapping current affects the histogram as $V_{sw}$ 
gets closer to $V_r$. The inset shows how the peak of the histogram moves with temperature.} 
\end{figure}

Figure \ref{fig:Hist} shows a selected number of histograms at different temperatures. The inset of 
Fig.\ \ref{fig:Hist} shows how the voltage value at the peak of the histogram changes versus 
temperature. At temperatures above about $130$mK, the displacement of the peak position is 
proportional to the temperature. We also see that the width of the distribution increases as the 
temperature is decreased. The retrapping voltage put a lower limit on the switching voltage and 
thereby sets a maximum temperature of about $180$mK. Therefore we limit ourselves to temperatures 
between $170$mK and the base temperature of the cryostat, about $50$mK. 

\begin{figure}[t]
\center 
\includegraphics[width=0.9\columnwidth]{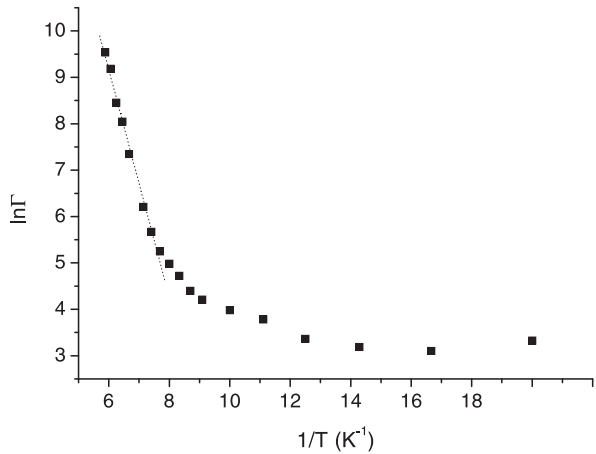} 
\caption{\label{fig:lnG}$\ln \Gamma$ versus the inverse temperature at $1.1$mV. Two temperature 
regimes are seen, the higher of which follows the thermal escape model.} 
\end{figure}

\begin{figure}[t]
\center 
\includegraphics[width=0.9 \columnwidth]{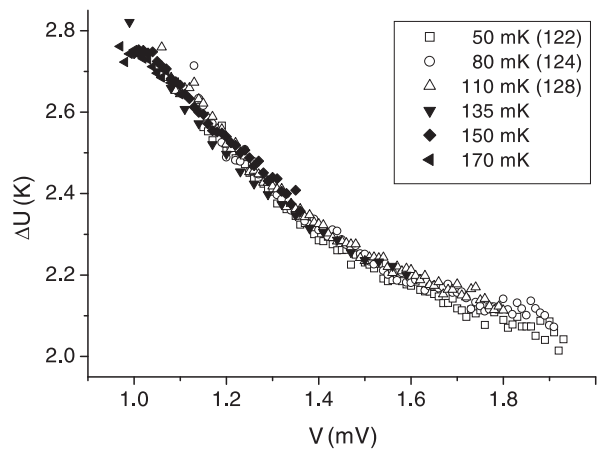} 
\caption{\label{fig:dUvsT}Selected curves showing how all data can be reduced to a single $\Delta 
U(V)$ curve in eq. (\ref{eq:ThermEsc}), with $\kappa \omega _p=80$GHz. The low temperature curves 
with the open symbols are fitted with an effective temperatures indicated in the parenthesis.} 
\end{figure}

From the histograms we can determine the escape rate $\Gamma(V)$ via equation (\ref{eq:CalcG}). In 
Fig.\ \ref{fig:lnG} we have plotted $\Gamma(V=1.1$mV) versus $1/T$. Here we see that in the high 
temperature region between $170$mK and $130$mK, the data is well described by Kramers' model of 
thermally activited escape, equation (\ref{eq:ThermEsc}). However at low temperatures, below ca 
$130$mK, the escape rate crosses over to a different behavior, becoming independent of temperature 
at the lowest temperatue attainable. By extrapolating the linear fit in the low temperature region 
shown in Fig.\ \ref{fig:lnG}, we find the intersection with the $\ln \Gamma$-axis and make an 
estimate of the factor, $\kappa \omega _p$ of the order of $10^9-10^{10}$Hz. An alternative way to 
find $\kappa \omega _p$, which uses the data at all voltages, is to extract the energy barrier 
$\Delta U(V)$ from the escape rate $\Gamma(V)$ via equation (\ref{eq:ThermEsc}). We then adjust 
$\kappa \omega _p$ so that the data at all temperatures fall on to a single $\Delta U(V)$ curve. We 
find that in the high temperature region between $170$mK and $130$mK, the data collapse on to one 
curve for $\kappa \omega _p=80$GHz as is shown in Fig.\ \ref{fig:dUvsT}.

We note that this method of determining $\kappa \omega _p$ differs from the traditional analysis of 
escape from the superconducting state of Josephson junctions. In the superconducting case we have a 
well established model for the energy barrier, and to an excellent approximation, escape from the 
cubic potential can be assumed. The cubic potential results in a straight line plot of 
$(\ln(\omega_p/2\pi\Gamma))^{2/3}$ versus the bias current \cite{martinis:phasedifference:87}. If 
we follow this analysis and plot $(\ln(\omega_p/2\pi\Gamma))^{2/3}$ versus bias voltage in the 
temperature range $130-170$mK using the value $\kappa \omega _p=80$GHz, we also find a straight 
line. However, the limited width of the distribution and the small temperature range make it 
difficult to say if the data follows the theoretical predictions (slope $\propto T^{-2/3}$ and 
intersect with the x-axis at $V_0$). We can, however, conclude that the qualitative behavior is 
correct.

\begin{figure}[b]
\center 
\includegraphics[width=0.9\columnwidth]{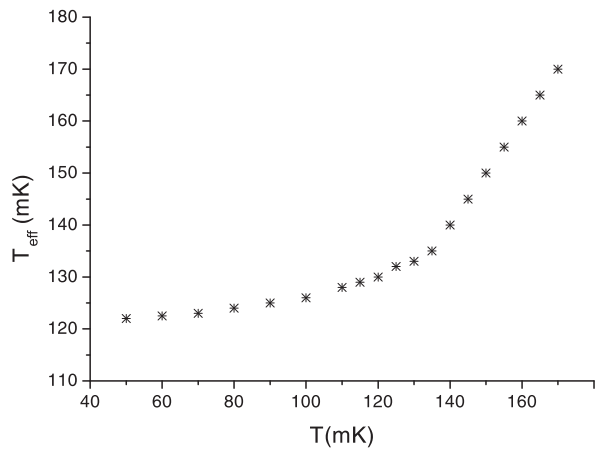} 
\caption{\label{fig:TeffVsT}The effective temperature, $T_{eff}$, is plotted versus the thermometer 
temperature, T. At high temperatures $T_{eff} \equiv T$. At low temperatures $T_{eff}$ was inferred 
by mapping all data to the same $\Delta U(V)$ curve.} 
\end{figure}

Thus the data are well described by the thermal activation model in the high temperature regime 
with $\kappa \omega _p=80$GHz. Fixing the value of $\kappa \omega _p=80$GHz, we can then analyze 
the low temperature regime ($T<130$mK) by adjusting the temperature in equation (\ref{eq:ThermEsc}) 
to find an effective temperature. Following this procedure, all data for all voltage and tempeature 
can be reduced to one $\Delta U(V)$ as shown for a few selected temperatures in Fig.\ 
\ref{fig:dUvsT}. The effective temperature, $T_{eff}$, is plotted versus the thermometer 
temperature in Fig.\ \ref{fig:TeffVsT}. Here we can see how the data can be described by the 
thermal escape model if an effective temperature is assumed that reaches a minimum value of $122$mK 
as the cryostat is cooled towards base temperature.

There can be several reasons for deviation from the thermal activation model at low temperatures. 
The escape temperature of $122$mK may be a result some source of heating of the sample, so that the 
sample is not in equilibrium with our thermometer. We can rule out self-heating by the current in 
the sample \cite{kautz:selfheating:93} because we are measuring escape from a zero current state, 
where there is no dissipation in the sample. The low temperature escape may also be influenced by 
RF and microwave noise reaching the sample through the filtered measurement leads. Another 
possibility is that the escape process at low temperatures is determined by macroscopic quantum 
tunneling (MQT). 

MQT rates have been calculated for the cubic potential at $T=0$ \cite{devoret:MQT:85},
\begin{equation}
  T_{esc} \approx {\frac{\hbar\omega _p/k_B}{7.2(1+0.87/Q)}}
\end{equation}
where Q is the damping factor. Assuming moderate to weak damping ($Q \gg 1$) and $\kappa \sim 1$ we 
can calculate an MQT escape temperature $T_{esc}\sim 85$mK, which is in reasonable agreement with 
the effective escape temperature found with our method of analysis. However, similar measurements 
and analysis on a slightly shorter array found a much lower $\kappa \omega _p \approx 10$MHz. This 
would result in $T_{esc}\sim 10 \mu$K in the MQT model, in poor agreement with the effective escape 
temperature $T_{eff}(0)\approx 100$mK found in the data analysis.

In conclusion, we have studied the onset of finite current in long, one-dimensional Josephson 
junction arrays in the Coulomb blockade state. By analyzing the temperature dependence of 
fluctuations of the switching voltage for the onset of finite current, we have shown that this 
onset can be described by the Kramers model for thermally activated escape over an energy barrier, 
where the barrier height depends on the bias voltage. The data fall into two temperature regimes.  
At high temperature thermal activation applies, but at low temperature an effective temperature 
must be assumed to fit the data to the Kramers model. The low temperature behavior could be the 
result of external noise causing the escape over the barrier, or possibly, macroscopic quantum 
tunneling through the barrier.

We would like to acknowledge helpful discussions with R.~Kautz and T.~Duty, and financial support 
from the NFR, G\"oran Gustafsson Foundation and the EU project SQUBIT.

\end{document}